\documentclass[preprint,prX]{revtex4}

\usepackage{graphicx}
\usepackage{tabularx}
\usepackage{booktabs}

\begin{document}


\title{Reflectance of Polytetrafluoroethylene (PTFE) for Xenon
  Scintillation Light}

\author{C. Silva}
\email{claudio@lipc.fis.uc.pt}
\author{J. Pinto da Cunha}
\author{A. Pereira, V. Chepel, M. I. Lopes,
  V. Solovov}
\author{F. Neves}
\affiliation{LIP-Coimbra
\\ Department of Physics, University of Coimbra
\\ P-3004 516 Coimbra, Portugal}
\date{\today}

\begin{abstract}

Gaseous and liquid xenon particle detectors are being used in a number of
applications including dark matter search and neutrino-less double beta
decay experiments. Polytetrafluoroethylene (PTFE) is
often used in these detectors both as electrical insulator and as a
light reflector to improve the efficiency of detection of scintillation
photons. However, xenon emits in the vacuum ultraviolet wavelength region
($\mathbf{\lambda}$=175 nm) where the reflecting properties of PTFE are not
sufficiently known.

In this work we report on measurements of PTFE reflectance,
including its angular distribution, for the xenon scintillation
light. Various samples of PTFE, manufactured by different processes
(extruded, expanded, skived and pressed) have been studied. The data were
interpreted with a physical model comprising both specular and diffuse
reflections. The reflectance
obtained for these samples ranges from about 
47\% to 66\% for VUV light.
Fluoropolymers, namely ETFE, FEP and PFA were also measured.

\end{abstract}

\pacs{29.40.Mc; 87.64.Cc; 78.20.Ci; 43.30.Hw; 42.25.Gy}

\keywords{Xenon scintillation; Reflectance; VUV; PTFE; Rough surfaces; Diffuse reflection}

\maketitle


\section{Introduction}

Liquid and gaseous xenon particle detectors are being used in a number of
applications including dark matter search \cite{zepii}, neutrino-less double beta
decay experiments \cite{dbb} and searches for $\mu\to e\gamma$ decays \cite{MEG}. 
Polytetrafluoroethylene (PTFE) is
often used in these detectors both as electrical insulator and
as a reflector to improve the detection efficiency of scintillation
photons. The reflectance distribution function of this material
  needs to be known in order to improve the simulations of the detectors
  and the data analysis of the light collection.
However, the xenon emits in the vacuum ultra violet region
($\lambda$=175 nm \cite{scin}) and, to our knowledge, the optical reflectance distribution 
of PTFE has not been measured
at those wavelengths. 

PTFE, also known as Teflon\textsuperscript{\textregistered}
($\mathrm{\left(C_{2}F_{4}\right)_{n}}$), is a polymer produced from tetrafluoroethylene. The
strong bound between carbon and
fluorine leads to a high chemical stability in a wide range of temperatures between
-200$^{\circ}$C and 260 $^{\circ}$C, which makes it suitable for use
in  xenon scintillation detectors.

In this work we report on measurements of PTFE reflectance,
including its angular distribution, for xenon scintillation
light. The data were
interpreted with a physical model comprising diffuse and specular lobes.

Measurements of the reflectance distributions are usually done inside a
gonioreflectometer illuminated by a laser beam, the reflected light being sampled at different angles.
However, not much data have been published on reflectance distributions, including data out of the
plane of incidence. Furthermore, most published data are for visible light, not for VUV
(\cite{PTFE}, \cite{ptfe2},\cite{bendler}, \cite{ptfe3}). 
At these wavelengths the measurement is challenging and must be
performed in vacuum or in a gas with low absorption in
the VUV.

We measured the reflectance of various PTFE samples, produced by different processes
(namely molded, skived, extruded and expanded). 
All these materials are obtained by suspension polymerization from tetrafluoroethylene, reduced to
a fine powder,
agglomerated in small pellets, and either: i) compressed in a mold, ii)
extruded, iii) skived or iv) expanded by air injection, respectively
\cite{synth}. 

PTFE belongs to a family of materials known as fluoropolymers. This
includes co-polymers of
tetrafluoroethylene and ethylene;
hexafluoropropylene; and $\mathrm{C_{2}H_{3}OC_{3}F_{7}}$, which are known as
ETFE (or Tefzel\textsuperscript{\textregistered}), FEP and PFA, respectively.
The reflectance of these materials was also measured and compared with
PTFE samples.

This work has been motivated by the need to derive a model of reflectance
to be included in simulations and data analysis
of detectors relying on production, propagation and 
detection of VUV light, namely those in which PTFE is used. We present
measurements of the reflectance distributions along with results of 
a physical reflectance model fitted to the data.

\section{Experimental Set-Up}

\begin{figure*}[!t]
\centering

\includegraphics[width=4.1in]{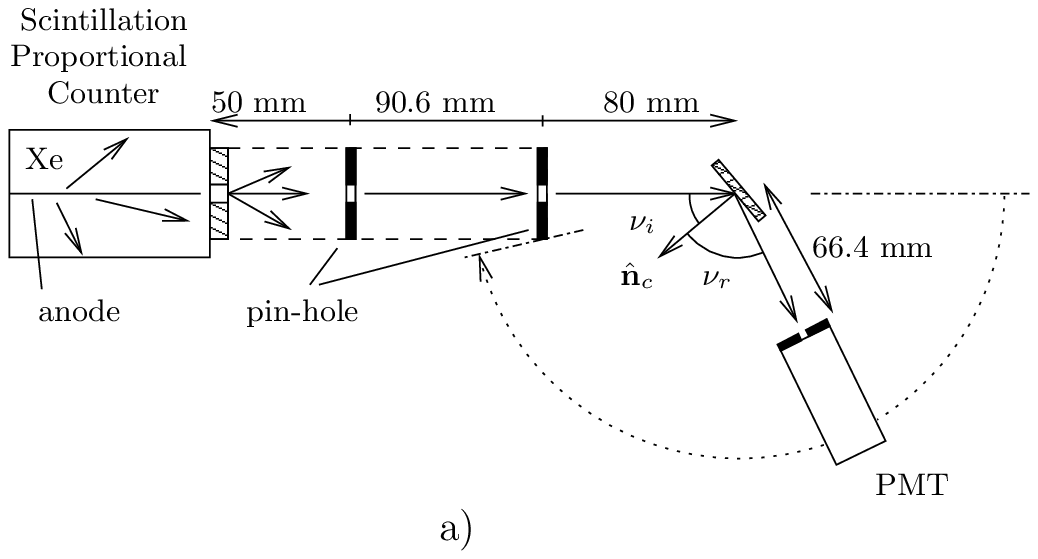}
\includegraphics[width=2.8in]{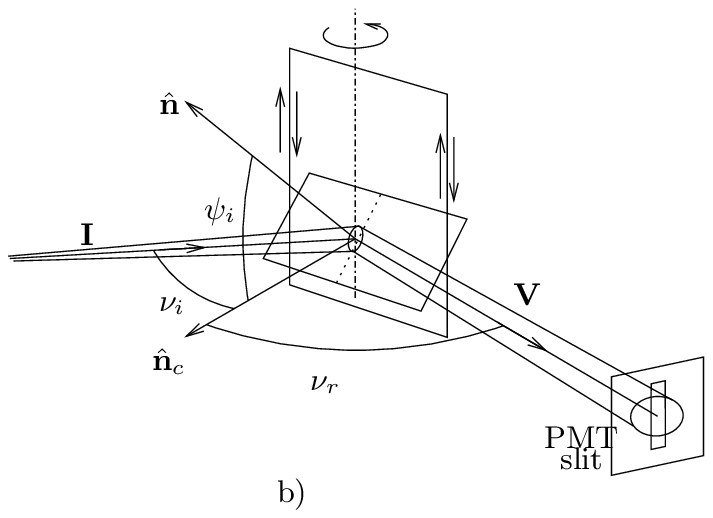}
\caption{The experimental set-up: a) top view and b) side view. 
The incident beam is along the direction ${\bf I}$ and 
${\bf V}$ points to the PMT position. The projection of the surface normal, ${\bf \hat{n}}$, in the plan of measurements
is ${\bf \hat{n}_{c}}$. The angles between ${\bf \hat{n}_{c}}$ and
$-{\bf I}$ and ${\bf \hat{n}_{c}}$ and ${\bf V}$ are $\nu_{i}$ and
$\nu_{r}$, as shown.
The inclination of the surface is given by the angle $\psi_i$.}
\label{front_view}
\end{figure*}

The experimental set-up used to measure the reflectance is depicted 
in Fig. \ref{front_view}, along with the system of coordinates.
The construction and characterization of the chamber is detailed elsewhere \cite{silva}.
The VUV light is produced  in a strong electric field near the central anode wire of a proportional
scintillation counter filled with gaseous xenon at 1.1
bar. Alpha-particles from $^{241}$Am are used as a source of
ionization. The light
exits through a fused silica window and is collimated by two pin-holes
before reaching the surface of the sample to be measured. 
The sample is rotated to change the angles $\nu_i$ and $\psi_i$ and thus the angle of incidence.
The light that is reflected
by the sample is detected by the photomultiplier (PMT) shown in Fig. \ref{front_view} at
different viewing angles, $\nu_r$.
The measurements are performed inside a stainless steel box filled with gaseous
argon, to prevent light absorption.


The reflected radiant intensity, $\frac{d\Phi_{r}}{d\Omega_{r}}$, is
  calculated dividing the number of photons detected in the direction
  $\nu_{r}$ by the solid angle comprised by the PMT window. To measure the
  incident beam flux, $\Phi_{i}$, we placed the PMT directly 
to the beam, having previously moved the sample off the beam course.

For each measurement we set $\nu_{i}$ and $\psi_{i}$ and measured the 
reflected light at different viewing direction angles, $\nu_{r}$.
The data taking time was in average $750\ $s, for each
point $\left\{\nu_{i},\psi_{i},\nu_{r}\right\}$. The background, measured by the PMT
(with beam on but no sample) was about 0.13 photoelectrons per second
, for all angles considered \cite{silva}.

The angles measured in the experiment: $\nu_i$, $\psi_i$ and $\nu_r$, are related with the 
angles $\theta_{i}$ and $\theta_{r}$ (Fig. \ref{sy_model}) used in
the reflection model through the following equations:
\begin{eqnarray}
\cos\theta_{i}& = &\cos{\nu_{i}}\cos{\psi_{i}}\nonumber\\
\cos\theta_{r}& = &\cos{\nu_{r}}\cos{\psi_{i}}
\end{eqnarray}

\section{Modelization of the reflection}

The light reflected from a surface is in general a superposition of diffuse
and specular contributions.
Its angular distribution depends of the 
structure of the surface, and of sub-surface inhomogeneities 
of the medium \cite{wolff}, which cannot be known in
detail. Therefore, 
the reflectance has to be treated statistically and parameterized to account for 
the roughness of the surface and the characteristics of the material.

Numerous approaches have been considered for describing the reflectance distributions.
These include both empirical and physically motivated models \cite{zhu}.
They often stem from the field of computer graphics, aiming at reproducing the
reflectance functions of real objects,
which is of paramount importance for perception in virtual reality environments. 
Though the focus is usually on visible light,
the goal is always to parametrize the reflectance in order
to reproduce the light distribution reflected from a surface.

\begin{figure}[!t]
\centering


\includegraphics[width= 3.5in]{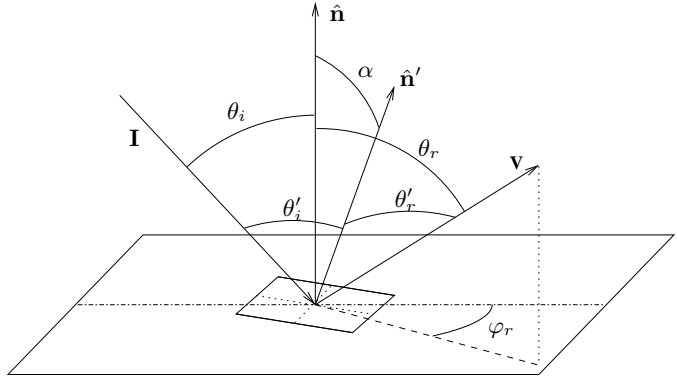}
\caption{The system of coordinates: ${\bf I}$ represents
  the direction of incidence of the photons, ${\bf V}$ is the viewing
  direction, and ${\mathbf{\hat{n}}}$ and ${\mathbf{\hat{n}}}'$ are surface 
  normal vectors, of the global (macroscopic) surface and of a local
  micro-surface, respectively.
 Primed angles are measured relatively to the local normal ${\mathbf{\hat{n}}}'$. }
\label{sy_model}
\end{figure}

The surface is modeled as an ensemble of micro-surfaces
  randomly oriented, according to some distribution function. 
Each micro-surface is defined by a normal $\mathbf{\hat{n}'}$, which is generally
non-coincident with the global (macroscopic) normal of the surface considered at large,
$\mathbf{\hat{n}}$.
Hence, two sets of angles have to be considered: i) global angles, relative 
to $\mathbf{\hat{n}}$, and ii) local angles, referred to
the normal of the local micro-surface,
$\mathbf{\hat{n}}'$ (see Fig. \ref{sy_model}).
As indicated, $\alpha$ is the angle between ${\mathbf{\hat{n}}}$ and
${\mathbf{\hat{n}'}}$. 
For specular reflection the relations between the
local and the global variables are
\begin{eqnarray}
\cos{2\theta_{i}'} &=&\cos{\theta_{i}}\cos{\theta_{r}}-\sin{\theta_{i}}\sin{\theta_{r}}\cos{\varphi_{r}}\\
\cos{\alpha} &=& \frac{\cos{\theta_{i}}+\cos{\theta_{r}}}{2\cos{\theta_{i}'}}
\end{eqnarray}

\begin{figure}[!t]
\centering

\includegraphics[width= 3.5in]{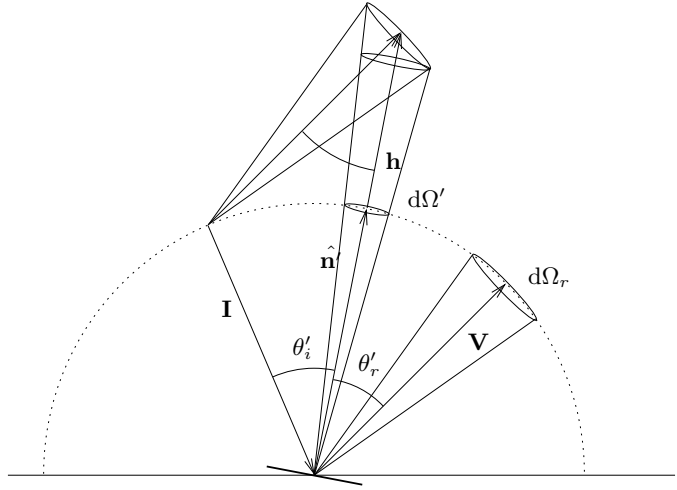}
\caption{In specular reflection, the micro-surfaces whose normals point
within a solid angle $d\Omega'$ radiate towards \textbf{V}, within the solid
angle $d\Omega_r$. Since
$d\Omega'={d\Omega_r\;\cos\theta'_i\over h^2}$, and
$h^2=(\mathbf{V}-\mathbf{I})^2$, consequently, $d\Omega_r=4\cos\theta'_i\,d\Omega'$.}
\label{f_solidangs}
\end{figure}

\begin{figure*}[!t]
\centering
\includegraphics[width=7.in]{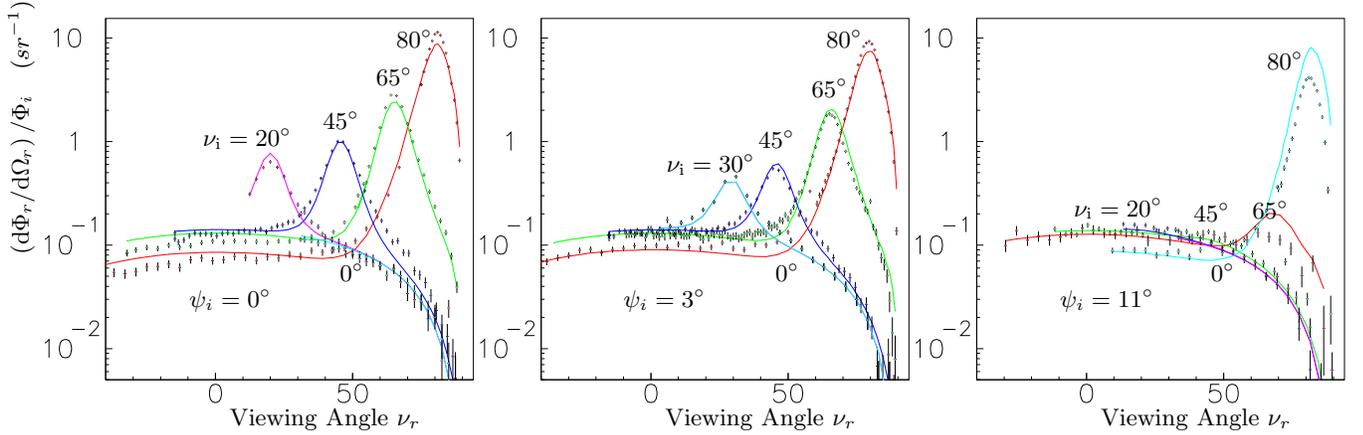}
\caption{The reflectance distribution of 
unpolished molded PTFE as a function of the viewing angle (in degrees), measured for three inclination angles $\psi_{i}$. 
The curves represent the predicted reflection upon an overall fit of the function 
$\varrho$  (eq. \ref{bridf}) to the entire data set (2223 data points in total), with three free
parameters.
The fitted
parameters have values: $n=1.51\pm0.07$, $\rho_L=0.52\pm0.06$, $\gamma = 0.057\pm0.008$.}
\label{ptfe_phi}
\end{figure*}

The diffuse radiance of a surface is often assumed to be independent
of the viewing direction (Lambert law),
but in fact it is non-Lambertian in general. 
This has been attributed to variations in the amount of light
that penetrates(exits) the sub-surface layers at different entry(exit) angles in result of the
Fresnel equations.
Hence, diffuse reflectance is best described by the Lambert law multiplied by a Fresnel coefficient
\cite{wolff}.
Moreover, a coarse surface is not Lambertian even if it is locally, 
since the radiance of several tilted Lambertian
micro-surfaces is not the same for all viewing directions \cite{oren}.
The resulting radiance is no longer Lambertian and, therefore, a
diffuse reflectance model has to encompass the two effects \cite{w-n-oren}.



The angular distribution of the reflectance can be conveniently expressed in terms of the
Bidirectional Reflected Intensity Distribution Function (BRIDF) \cite{nicodemus}, 
which is by definition
the intensity of the reflected light into a given direction per unit of incident beam 
flux,
\begin{equation}
\varrho(\theta_i,\varphi_i,\theta_r,\varphi_r)
={d\Phi_r/d\Omega_r\over\Phi_i}
\end{equation}
where $d\Omega_r$ is an element of the solid angle towards the viewing direction 
and $\Phi_i$ and $\Phi_r$ are the incident and reflected fluxes of radiation, respectively.
This function 
describes how reflected light is distributed in space for any given direction of incidence.
It is therefore a suitable quantity to parametrize the observations.
Given that the reflection has specular and diffuse components it is
convenient to distinguish their
contributions in the function $\varrho$,
$$
\varrho=\varrho_S+\varrho_D
$$

The incident flux of radiation impinging at a certain area $\delta A$ of the surface is 
$$
d\Phi_i=L_i \cos\theta_i\,\delta A\,d\Omega_i
$$
where $L_i$ is the radiance of the source and $d\Omega_i$ is the solid angle subtended by the
incident beam.
The surface is \textit{a priori}
composed by many micro-surfaces 
and each micro-surface of $\delta A$ has its own normal $\mathbf{\hat{n}'}$, distributed
according to some probability distribution function, $P$.
The number of normals $\mathbf{\hat{n}'}$
pointing within
a solid angle $d\Omega'$ is $P\,d\Omega'$
and the effective area of the micro-surfaces whose normal is within the solid angle $d\Omega'$ is
$Pd\Omega'\,\delta A$.
We assume that the micro-surfaces have no preferred direction and, therefore, $P=P(\alpha)$. The incident flux at the micro-surfaces is
$$
d\Phi'_i=L_i\cos\theta'_i\,\delta A\,d\Omega_i P d\Omega'
$$
Therefore, the specular radiated flux by the area $\delta A$ into a
direction $\hat{{\bf k}}_r$ is
given by
\begin{equation} 
d\Phi_r^{(S)}=FG\,d\Phi'_i=FG\,d\Phi_i
{\cos\theta'_i\over\cos\theta_i}P d\Omega'
\end{equation}
where the geometrical attenuation factor $G$ 
accounts for shadowing and masking between micro-surfaces \cite{smith}
and the Fresnel coefficient, $F$, expresses the fraction of light that
is reflected at the surface with normal $\mathbf{\hat{n}'}$.
For unpolarized light, 
\begin{equation}
F(\theta_i,n/n_0)={1\over2}
{\sin^2(\theta_i-\theta_t)\over\sin^2(\theta_i+\theta_t)}
\left[1+{\cos^2(\theta_i+\theta_t)\over\cos^2(\theta_i-\theta_t)}\right]
\end{equation}
where $\theta_t=\arcsin\left(n_0/n\,\sin\theta_i\right)$, and 
$n$ and $n_0$ are the refraction indices above and bellow the surface,
respectively.  The angles are defined in Fig.\ref{sy_model}.

For specular reflection the following relation holds \cite{nayer}, \cite{walter} (see Fig.
\ref{f_solidangs})
\begin{equation}
d\Omega_r=4\cos\theta'_i\,d\Omega'
\end{equation}
and, therefore,
\begin{equation}
\varrho_S={FGP\over4\cos\theta_i}
\end{equation}
This argument follows closely the work of Torrance-Sparrow \cite{sparrow}.

The diffuse reflection flux by a smooth surface is given in Lambert approximation by
\begin{equation}
d\Phi_r^{(D)}={\rho_L\over\pi}W\,d\Phi_i \cos\theta_r\,d\Omega_r
\end{equation}
where $\rho_L$ is the diffuse albedo, and the factor 
\mbox{$W=[1-F(\theta_i,n/n_0)][1-F\left(\arcsin\left[n_0/n\,\sin\theta_r,n_0/n\right]\right)]$} 
is the Wolff correction factor that accounts for entry(exit) of light into(out of)
the sub-surface layer \cite{wolff}. 
If the roughness of the surface is taken into account and 
a distribution of micro-surfaces is considered, then the 
diffuse reflection flux from a distribution of micro-surfaces ought to be
$$
d\Phi_r^{(D)}={\rho_L\over\pi}\,{d\Phi_i\over\cos\theta_i}d\Omega_r
\int W\cos\theta'_i\,\cos\theta'_r\,G{Pd\Omega'\over\cos\alpha}
$$
where $P d\Omega'$ is the fraction of $\delta A$ whose normal is within  $d\Omega'$ and
the factor ${1\over\cos\alpha}$ accounts for the effectively radiating micro-area.
Hence 
\begin{equation}
\varrho_D\approx{\rho_L\over\pi\cos\theta_i}\,W\cos\theta_r
\int {\cos\theta'_i\,\cos\theta'_r\over\cos\theta_r\cos\alpha}GPd\Omega'
\end{equation}
This model proved successful in describing the diffuse reflection of various samples illuminated
with visible light \cite{w-n-oren}. The integral has been parametrized numerically 
in the form
\begin{equation}
\left(1-{\cal A}+{\cal B}\right)\cos\theta_i
\end{equation}
where
\begin{eqnarray}
{\cal A}&\simeq&0.5 \frac{\gamma^{2}}{\gamma^{2}+0.92}\nonumber \\
{\cal B}&\simeq&0.45\frac{\gamma^{2}}{\gamma^{2}+0.25}
H(\cos\varphi_r)\cos\varphi_r\sin\theta_M\tan\theta_m\nonumber
\end{eqnarray}
and $\gamma$ is the width of the distribution $P=P(\alpha)$ considered
(eq. \ref{a3});
$H(x)$ is the Heaviside step function; \mbox{$\theta_m=\mbox{min}(\theta_i,\theta_r)$}; and
\mbox{$\theta_M=\mbox{max}(\theta_i,\theta_r)$}. 
Note that if $\gamma=0$ this equation reduces to the Lambert law. 

Putting it altogether, the reflectance distribution is described by the function
\begin{equation}
\varrho(\theta_i,\varphi_i,\theta_r,\varphi_r)=
{FGP\over4\cos\theta_i}
+{\rho_L\over\pi}
W\left(1-{\cal A}+{\cal B}\right)\cos{\theta_{r}}
\label{bridf}
\end{equation}



We considered that the surface is invariant to rotations about the normal,
in which case $\varphi_i$ can be set to zero.
The function $\varrho$ above 
has three parameters which have to be evaluated from observation of the light
reflected by the surface: $\rho_L$, $n$ and $\gamma$. 
These coefficients are in general functions of the wavelength, and additionally $\rho_L<1$ 
to be consistent with flux conservation.


The first term of function $\varrho$ represents the specular
reflection lobe \cite{sparrow}, whose width is proportional
to the roughness of the surface, embodied in the distribution $P$. 
For the distribution function $P$ we considered the function that was deduced by
Trowbridge and Reitz assuming for the micro-surfaces an ellipsoidal shape \cite{reitz},
\begin{equation}
  P\left(\alpha;\gamma\right) =
  \frac{\gamma^{2}}{\pi\cos^{4}{\alpha}\left(\gamma^{2}+\tan^{2}\alpha\right)^{2}}\label{a3}\\
\end{equation}
In fact, this distribution seems the most successful in describing the
data, when compared with other forms
\cite{walter}.
As for the geometric attenuation factor $G$,
we considered the closed form due to Smith \cite{smith}:
\begin{equation}
G(\theta_i,\theta_r,\varphi_r)\simeq 
H(\theta'_i-\pi/2)H(\theta'_r-\pi/2)G'(\theta_i)G'(\theta_r)
\end{equation}
where
\begin{equation}
G'(\theta;\gamma)={2\over1+\sqrt{1+\gamma^{2}\tan^{2}\theta}}
\end{equation}
Though, this factor is only relevant for angles above $80^\circ$.
%

Our goal is to fit the measured data with a function having the minimum 
physically meaningful parameters. The above model has only three free
parameters, i.e. n, $\rho_{L}$ and $\gamma$, whose values
are determined from a fit to the data set of points measured.
A least squares minimization method based on a genetic algorithm was implemented
specifically for this purpose.
In evaluating $\varrho$, it is necessary to set the variable $\alpha$ 
by stochastic sampling according to the probability distribution function $P(\alpha;\gamma)$
referred above.
The fit results are shown in the table \ref{par_val}. Examples of
the fits are shown in figures \ref{ptfe_phi} to \ref{flur}.

The total reflectance is obtained by numerical integration of $\varrho$
over all viewing directions, 
\begin{equation}
R(\theta_i,\varphi_i)=\int
\frac{1}{G}\,\varrho(\theta_i,\varphi_i,\theta_r,\varphi_r)\sin{\theta_{r}}d\theta_rd\varphi_r
\end{equation}
using the fitted parameters mentioned above. The factor $1/G$ recalls
that the integral includes all reflected light irrespective of its direction.
This integral has been calculated numerically for
    the diffuse and the specular components of reflection. 
    The results are plotted in Fig. \ref{ptfe_totr}
and shown in table \ref{manaf_int} for various surfaces. 

The fitted function $\varrho(\theta_i,\phi_{i},\theta_r,\varphi_r)$
stated above
can be readily implemented in the Monte Carlo simulation of the 
propagation of light in scintillation detectors.


\section{Experimental Results}

We measured the reflectance distribution of a $5\ $mm tick
piece of unpolished molded-PTFE for angles of incidence
\mbox{$\nu_{i}\in\left\{0^{\circ},20^{\circ},30^{\circ},45^{\circ},55^{\circ},65^{\circ},
80^{\circ}\right\}$} and inclinations \mbox{$\psi_{i}\in\left\{0^{\circ}, 3^{\circ},
8^{\circ}, 11^{\circ}, 20^{\circ}\right\}$}.

The BRIDF function $\varrho$ was fitted to the entire data set (2223
points), yielding for unpolished 
molded-PTFE:
$n=1.51\pm0.07$, $\rho_L=0.52\pm0.06$ and $\gamma=0.057\pm0.08$.
A subset of these results is shown in Fig.
\ref{ptfe_phi}, for the angles indicated.
The curves represent the reflectances predicted by the overall fit to all data points measured,
including measurements out of the incidence plan.
This simple model seems to reproduce the main features observed in the data,
despite the fact that data at high angles are included in the fit.


These results show that for $\psi_{i}\ge10^{\circ}$ the specular lobe
is highly suppressed, whereas
the intensity of the diffuse component does not
change significantly, as would be expected for a consistent data set.


\begin{figure*}[!t]
\centering
\includegraphics[height=2.0in]{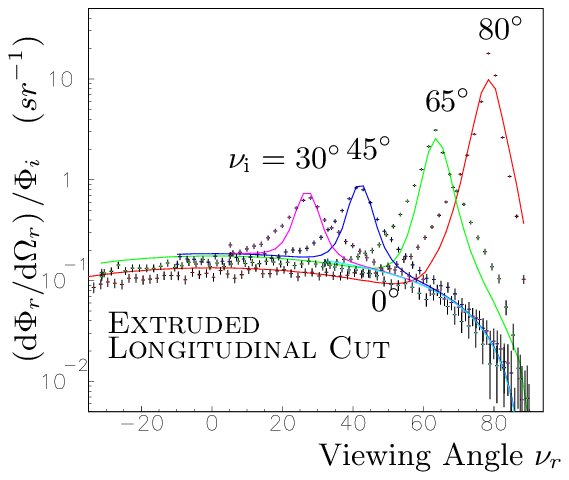}
\includegraphics[height=2.0in]{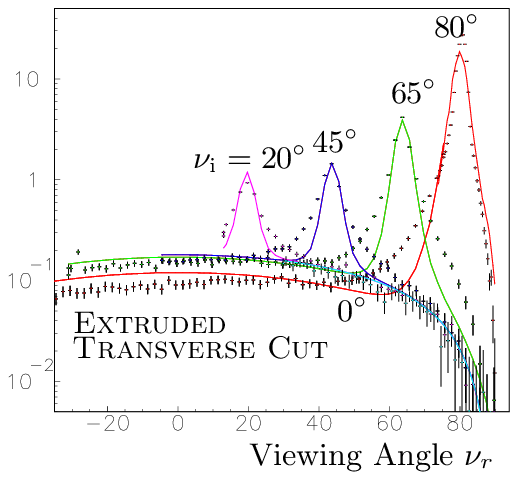}
\includegraphics[height=2.0in]{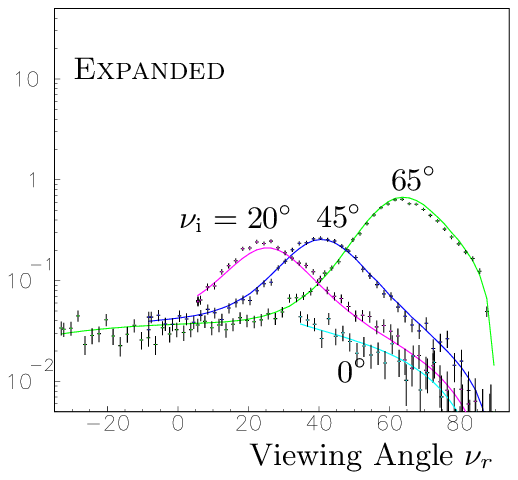}

\includegraphics[height=2.0in]{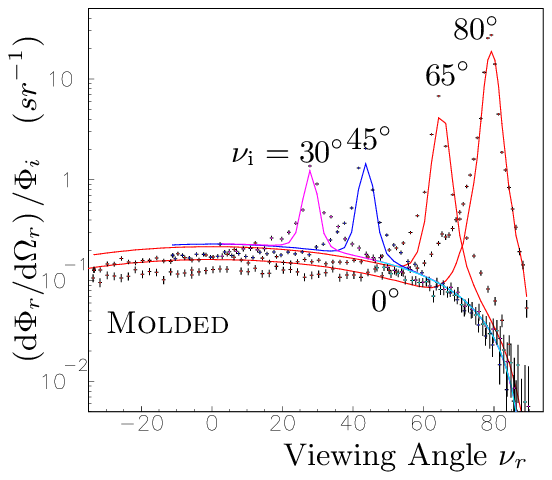}
\includegraphics[height=2.0in]{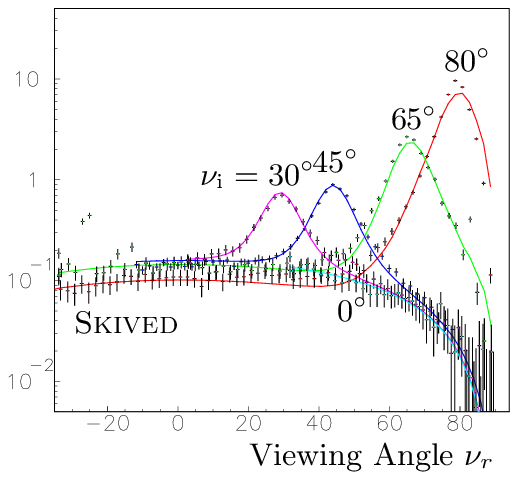}
\caption{Reflectance distributions of various PTFE samples, produced as indicated, plotted as a
function of the viewing angle (in degrees), for various angles of incidence.
}
\label{manufacture}
\end{figure*}

\begin{figure}
\centering
\includegraphics[width=3.0in]{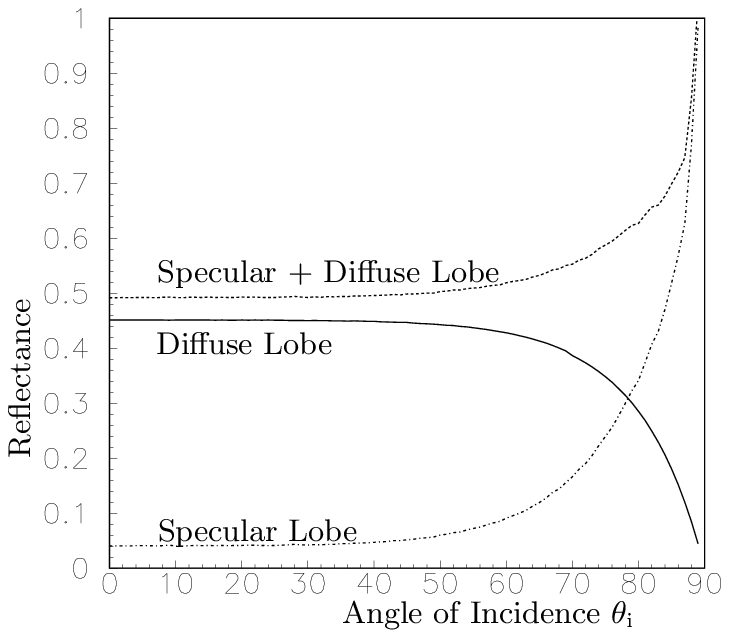}
\caption{The reflectance of the unpolished molded PTFE sample as a function of the angle of
  incidence $\theta_{i}$ (in degrees), for light of $\lambda$=175 nm. }
\label{ptfe_totr}
\centering
\includegraphics[width=3.0in]{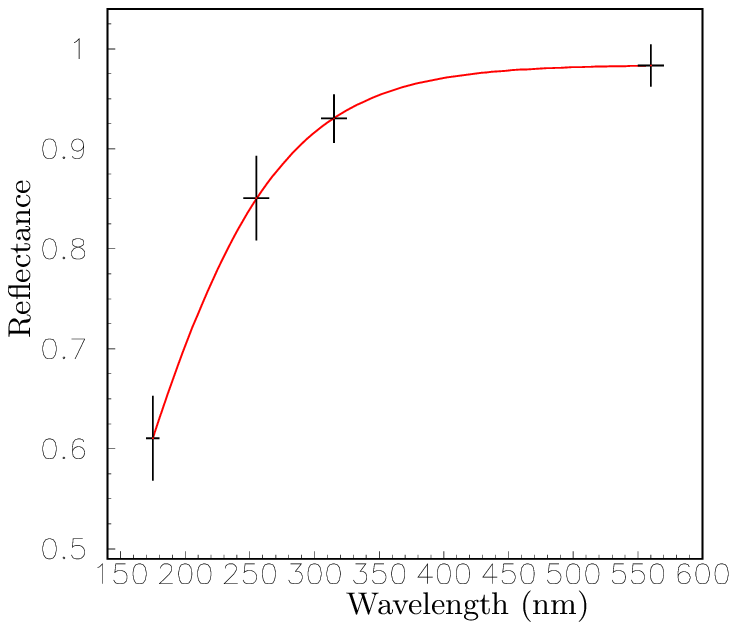}
\caption{Reflectance of the molded unpolished PTFE sample as a function of the wavelength, at
  the normal incidence ($\theta_{i}=0$).}
\label{wavelength}
\end{figure}

The integral of the reflectance function $\varrho$ (eq. \ref{bridf})
is represented in Fig. \ref{ptfe_totr} as a function of the angle of incidence,
for specular and diffuse reflection components.
The results show that the intensity of the diffuse lobe is nearly constant
up to about $60^{\circ}$ ,
whereas the specular lobe increases with increasing angle of incidence
as would be expected from the Fresnel equations. The specular lobe
becomes more intense than the diffuse lobe for
$\theta_{i}\gtrsim79^{\circ}$.


In Fig. \ref{manufacture} we show reflectance measurements of different kinds of PTFE,
manufactured by the processes indicated, for light of $175\,$nm.
These samples had thickness of $7\ $mm ($5\ $ mm for extruded-PTFE)
and were finished with  ultra fine sandpaper (P2000) and polished prior to measurement
(except expanded-PTFE owing to its mechanical properties).
Two samples of extruded-PTFE were measure, cut along and
transversely to the direction of extrusion, respectively (as they might have a preferred direction).
The surface inclination angle was set to $\psi_{i}=0^{\circ}$ in all cases represented in Fig. \ref{manufacture}.

\begin{table*}[!t]
\caption{Values of $n$, $\rho_L$ and $\gamma$ obtained from a fit to the measured data, 
for the samples indicated. The
"Extruded${}_{\perp}$ and "Extruded${}_{\parallel}$ refer to
cuts perpendicular and parallel to the extrusion direction.
All PTFE samples (but expanded-PTFE) were polished prior to
measurement. Systematic and statistical uncertainties are added in quadrature.}
\centering
\begin{tabular}{llcccc}
\hline
Fluoropolymer & & $n$     & $\rho_L$  &
$\gamma$ \\
\hline
PTFE&Molded Unpolished     &$1.51\pm0.07$  &$0.52\pm0.06$   &$0.057\pm0.008$ \\
PTFE&Molded Polished       &$1.30\pm0.09$  &$0.82\pm0.09$   &$0.014\pm0.005$ \\
PTFE&Extruded${}_\perp$     &$1.35\pm0.03$&$0.73\pm0.07$  &$0.019\pm0.010$ \\
PTFE&Extruded${}_\parallel$ &$1.32\pm0.06$  &$0.65\pm0.04$   &$0.033\pm0.012$          \\
PTFE&Skived                &$1.49\pm0.07$  &$0.580\pm0.013$ &$0.064\pm0.006$   \\
PTFE&Expanded              &$1.56\pm0.05$ &$0.14\pm0.03$ &$0.146\pm0.011$     \\
PFA &                      &$1.30\pm0.06$  &$0.69\pm0.05$   &$0.012\pm0.007$\\
FEP &                      &$1.25\pm0.10$  &$0.24\pm0.02$ &$0.0092\pm0.015$\\
ETFE&                      &$1.33\pm0.12$  &$0.07\pm0.07$   &$0.007\pm0.003$\\
\hline
\end{tabular}
\label{par_val}
\end{table*}

The curves in Fig. \ref{manufacture} represent the best overall fit of function $\varrho$,
considering all data.
The values of the fitted parameters are shown in table \ref{par_val}. As
shown, the index of refraction appears to be between 1.25 for the
FEP sample and 1.56 for the expanded sample. The
width of the specular lobe is between 0.014 and 0.064 for the molded
and skived surfaces, respectively.

The reflectance distributions are similar for extruded, skived and
molded samples. However the reflection of expanded-PTFE appears to be quite
different from the others, showing less diffuse light and a much
broader specular lobe.

\begin{table*}[!t]

\caption{Reflectance of various types of PTFE, manufactured as indicated, 
for two angles of incidence $\theta=0^\circ$ and $\theta=65^\circ$.
Extruded${}_\perp$ and Extruded${}_\parallel$ refer to surfaces 
cut perpendicular and parallel to the extrusion direction, respectively.
}
\centering
\begin{tabular}{lccccccc}
\hline


& \multicolumn{3}{c}{{
    $0^{\circ}$}}&&\multicolumn{3}{c}{{$65^{\circ}$}}\\

\hline
&$\mathrm{R_{{diffuse}}}$ & $\mathrm{R_{specular}}$ &
$\mathrm{R_{total}}$&$\ \ $ & $\mathrm{R_{diffuse}}$ & $\mathrm{R_{specular}}$& $\mathrm{R_{total}}$ \\
\hline
Molded Unpolished     & $0.45\pm0.02$ & $0.041\pm0.009$&$0.49\pm0.02$ &&$0.41\pm0.02$  &$0.115\pm0.012$&$0.53\pm0.02$\\ 
Molded Polished       & $0.72\pm0.06$ & $0.019\pm0.006$&$0.74\pm0.06$ &&$0.73\pm0.08$  &$0.08\pm0.02$ &$0.81\pm0.08$\\
Extruded${}_{\perp}$ 
                      & $0.67\pm0.06$ & $0.022\pm0.003$&$0.69\pm0.06$ &&$0.64\pm0.04$  &$0.090\pm0.007$ &$0.73\pm0.04$\\
Extruded${}_{\parallel}$  & $0.59\pm0.04$ & $0.019\pm0.006$&$0.61\pm0.04$ &&$0.56\pm0.05$  &$0.0825\pm0.012$&$0.64\pm0.05$\\
Skived                & $0.51\pm0.02$ & $0.039\pm0.008$&$0.55\pm0.02$ &&$0.47\pm0.02$&$0.12\pm0.02$ &$0.59\pm0.02$\\
Expanded              & $0.11\pm0.03$ & $0.048\pm0.007$&$0.16\pm0.03$ &&$0.10\pm0.03$&$0.111\pm0.009$ &$0.21\pm0.03$\\
PFA                   & $0.63\pm0.07$ & $0.017\pm0.006$&$0.65\pm0.07$ &&$0.58\pm0.03$&$0.08\pm0.02$ &$0.66\pm0.04$\\
FEP                   & $0.23\pm0.02$ & $0.012\pm0.009$&$0.24\pm0.02$&&$0.22\pm0.02$&$0.04\pm0.02$ &$0.25\pm0.03$\\
ETFE                  & $0.065\pm0.03$& $0.020\pm0.012$&$0.085\pm0.012$&&$0.10\pm0.07$&$0.09\pm0.03$&$0.19\pm0.07$\\
\hline
\end{tabular}

\label{manaf_int}
\end{table*}

The reflectance of each sample is obtained by integrating the distribution function $\varrho$
for the fitted
parameters referred above. The results are presented 
in table \ref{manaf_int} for angles of incidence
$\theta_{i}=0^{\circ}$ and $\theta_{i}=65^{\circ}$. They show that the reflectance is
mainly diffuse, with the diffuse lobe accounting for more than 90\% of the reflection
at $\theta_{i}=0^{\circ}$ (70\% for expanded-PTFE).
At $\theta_{i}=$65$^{\circ}$ the diffuse lobe accounts for more than
80\% of the reflection (48\% for the expanded sample).
These results show clearly that the relative contribution of the
 diffuse lobe is larger for the samples with higher reflectance.

The expanded PTFE  shows the lowest reflectance of all measured.
The fact that the specular component does not follow this trend 
might suggest that VUV light is being absorbed by oxygen molecules 
trapped in the pores of the material, underneath the surface \cite{huang}.

It can be concluded from these results that the reflectance of
polished-molded-PTFE is about 74\% (table \ref{manaf_int}), whereas the corresponding 
unpolished sample has a reflectance of only 49\% (see Fig. \ref{ptfe_totr}).
Polishing the surface enhances the total reflectance of PTFE.
This might be due to
air trapped in the porous PTFE that  is released by the polishing, but
further studies are necessary to fully understand this effect.

\begin{figure*}[!t]
\includegraphics[height=2.0in]{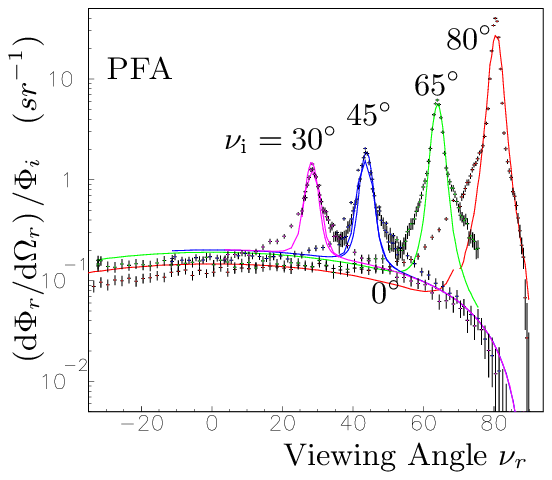}
\includegraphics[height=2.0in]{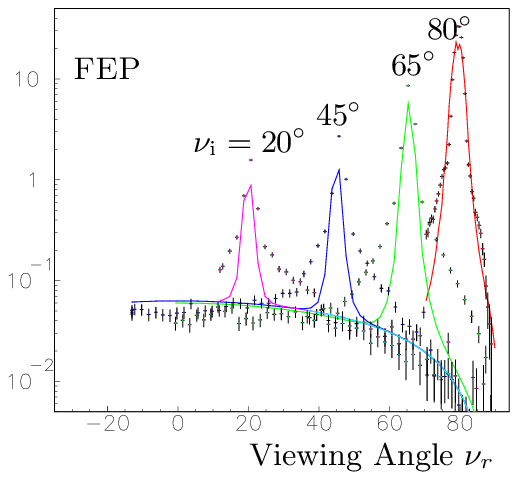}
\includegraphics[height=2.0in]{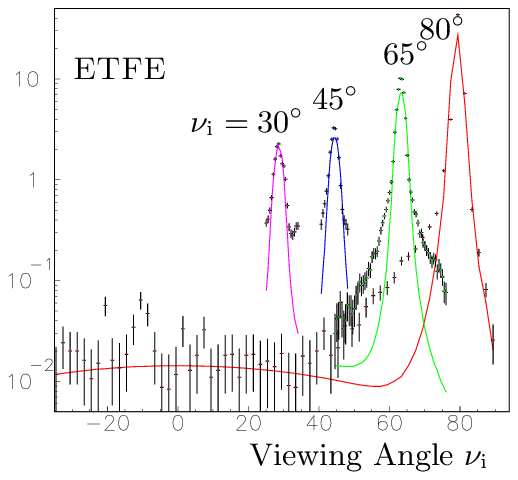}
\caption{Reflectance distribution 
as a function of the angle of incidence (in degrees), for
the fluoropolymers indicated.}

\label{flur}
\end{figure*}

The observed reflectance of 
fluoropolymers: ETFE, FEP and PFA is shown in Fig. \ref{flur}.
Samples of thickness 0.5 mm (ETFE), 1.1 mm
(FEP) and 1.5 mm (PFA) were measured.
All of them are transparent to visible light, but proved
to be opaque at 175 nm, as concluded from the fact that
no light was detected by the PMT on the other side of the samples illuminated with VUV light.

As shown in  Fig. \ref{flur}, PFA and FEP have both diffuse and specular reflection  components which are 
similar to those of PTFE,
for light of 175 nm wavelength. The values of the fitted parameters are  included in table
\ref{par_val}.
However, 
ETFE shows only a
specular lobe, the diffuse component being much suppressed, compared
to the former. 

The model considered above fails to fully describe the tails of the
samples with 
narrow lobes, notably the fluoropolymers ETFE, FEP and PFA, which
might be indicating that the contribution from coherent reflection should also
be considered in these cases. 
Hence, the indices of
refraction of these fluoropolymers are possibly under-estimated (table \ref{par_val}).

The reflectance of molded-PTFE was also studied as a function of wavelength
using light emitting diodes (LED), emitting at 
\mbox{$\lambda=\{250; 300; 550\}\ $nm}, respectively.
These results are shown in Fig. \ref{wavelength} and in table
\ref{total_manaf} for light at normal incidence. The index of
refraction of PTFE obtained at  550 nm (1.361) agrees
with published values (between 1.3 and 1.4 \cite{ferry}). The reflectance increases with the wavelength, 
reaching a value about $98\%$ for light of $550\ $nm, in line with
previous observations of PTFE at the near ultraviolet \cite{PTFE}.
This is due to the increase of the diffuse component due to 
internal scattering with increasing wavelength, while
the specular lobe remains fairly constant above 250 nm. Thus, PTFE is
a good diffuser for visible
light, the specular reflection being only about $3.4\%$ at normal incidence 
for a wavelength of the order of $\lambda=550\,$nm. However, these results show that 
the specular reflection cannot be ignored for VUV light, owing to the
decrease of the diffuse component.

\begin{table}[!t]
\caption{Fitted values of $n$, $\rho_L$ and $\gamma$ for the unpolished molded PTFE 
  sample at the
wavelengths measured.}
\begin{tabular}{cccc}
\hline
Wavelength & $n$     & $\rho_L$ &$\gamma$ \\
(nm)       &&&\\
\hline
175&$1.51\pm0.07$&$0.52\pm0.06$&$0.057\pm0.008$ \\
250&$1.31\pm0.03$&$0.91\pm0.04$&$0.059\pm0.007$          \\
310&$1.31\pm0.02$&$0.99\pm0.02$&$0.049\pm0.004$ \\
550&$1.36\pm0.02$&$1.06\pm0.02$&$0.0414\pm0.0006$    \\
\hline
\end{tabular}
\label{total_manaf}
\end{table}

\section{Conclusions}

The reflectance distribution of PTFE at 175 nm clearly shows specular
and diffuse lobes.
At these wavelengths PTFE cannot be considered a "perfect" diffuser.
The amount of specularly reflected light can be significant in scintillation detectors using this material.
The experimental observations were interpreted using a
function distribution which models the diffuse and specular reflections by rough
surfaces. The data are well described by a generic BRIDF function, $\varrho$, with only three free
parameters. The fitted $\varrho$ can be used to obtain the
reflectance of the considered surfaces of PTFE, by integration over all viewing angles.

The PTFE samples that were polished show a more narrow specular lobe and a
more intense diffuse
lobe compared to the unpolished surfaces. Moreover, the reflectance distribution is similar for the
various PTFE samples observed, with the notable exception of
expanded-PTFE. The reflectance obtained, at normal incidence
varies between 44\% and 66\% (is about 15\% for expanded-PTFE), at $\lambda=175\,$nm.

In all cases the diffuse lobe is the dominant reflection component of PTFE.
It accounts more than 90\%  of the total reflectance at
$\theta_{i}=0^{\circ}$ (70\% for expanded-PTFE). It was also found
that the reflectance of the PTFE increases with increasing
wavelength, due to the growth of the diffuse reflection.
Thus, the contribution of the specular lobe to the total reflectance increases at
  lower wavelengths.

The reflectances of PFA and FEP resemble that of PTFE, even if
FEP has a less intense diffuse component
compared to PTFE. By contrast, the ETFE surface is mostly specular for VUV light of 175 nm.

The reflection is reasonably described by the distribution function $\varrho$ discussed above.
Having only three parameters this function is suitable to
parameterize the reflection by 
PTFE surfaces illuminated with VUV light, especially in the context of simulating the
light collection in particle physics scintillation detectors.

Finally it would be interesting to investigate if the reflectance distribution of PTFE 
in contact with gas is dramatically altered if
the surface is immersed in liquid xenon, more than would be expected from changing the refraction 
index. The results depend on
whether the liquid enters into the any superficial pores.

\vfill

\section*{Acknowledgment}

The samples of FEP, ETFE and PFA were kindly offered by DuPont de Nemours.
This work is supported by the FCT projects 
POCI/FP/81928/2007 and CERN/FP/83501/2008, 
C. Silva was supported by the FCT fellowships SFRH/BD/19036/2004.


%

\bibliography{paper_claudio_APS_e}

\end{document}